\documentclass[12pt]{article}
\usepackage{a4wide}
\usepackage{amssymb, amsmath,cite}
\usepackage{hyperref}
\usepackage[normalem]{ulem}
\usepackage[utf8]{inputenc}
\usepackage[usenames,dvipsnames]{color}

\begin{document}
{\renewcommand{\thefootnote}{\fnsymbol{footnote}}
		
\begin{center}
{\LARGE Swampland, Trans-Planckian Censorship and Fine-Tuning Problem for Inflation:\\[1mm] Tunnelling Wavefunction to the Rescue} 
\vspace{1.5em}

Suddhasattwa Brahma$^{1}$\footnote{e-mail address: {\tt suddhasattwa.brahma@gmail.com}}, Robert Brandenberger$^{1}$\footnote{e-mail address: {\tt rhb@physics.mcgill.ca}} and Dong-han Yeom$^{2,3}$\footnote{e-mail address: {\tt innocent.yeom@gmail.com}}  
\\
\vspace{0.5em}
$^{1}$ Department of Physics, McGill University,\\
 Montr\'eal, QC H3A 2T8, Canada\\
 
$^{2}$ Department of Physics Education, Pusan National University, \\
 Busan 46241, Republic of Korea\\
 
$^{3}$ Research Center for Dielectric and Advanced Matter Physics, \\
Pusan National University, Busan 46241, Republic of Korea 
\vspace{1.5em}
\end{center}
}
	
\setcounter{footnote}{0}

\newcommand{\bea}{\begin{eqnarray}}
\newcommand{\eea}{\end{eqnarray}}
\renewcommand{\d}{{\mathrm{d}}}
\renewcommand{\[}{\left[}
\renewcommand{\]}{\right]}
\renewcommand{\(}{\left(}
\renewcommand{\)}{\right)}
\newcommand{\nn}{\nonumber}
\newcommand{\Mpl}{M_{\textrm{Pl}}}
\newcommand{\Hinf}{H_{\textrm{inf}}}
\def\H{\mathrm{H}}
\def\V{\mathrm{V}}
\def\e{\mathrm{e}}
\def\be{\begin{equation}}
\def\ee{\end{equation}}

\begin{abstract}

\noindent The trans-Planckian censorship conjecture implies that single-field models of inflation require an extreme fine-tuning of the initial conditions due to the very low-scale of inflation. In this work, we show how a quantum cosmological proposal -- namely the tunneling wavefunction -- naturally provides the necessary initial conditions without requiring such fine-tunings. More generally, we show how the tunneling wavefunction can provide suitable initial conditions for hilltop inflation models, the latter being typically preferred by the swampland constraints.

\end{abstract}

\section{TCC Bounds on Inflation}

The trans-Planckian censorship conjecture (TCC), proposed recently \cite{TCC}, aims to resolve an old problem for inflationary cosmology. The idea that observable classical inhomogeneities are sourced by vacuum quantum fluctuations \cite{ChibMukh, Starob} lies at the heart of the remarkable success of inflationary predictions. On the other hand, if inflation lasted for a long time, then one can sufficiently blue-shift such macroscopic perturbations such that they end up as quantum fluctuations on trans-Planckian scales \cite{Jerome}. This would necessarily require the validity of inflation, as an effective field theory (EFT) on curved spacetimes, beyond the Planck scale. This is what the TCC aims to prevent by banishing \textit{any} trans-Planckian mode from ever crossing the Hubble horizon and, thereby, setting an upper limit on the duration of inflation. Yet, in order to explain current horizon scale inhomogeneities as being sourced by vacuum quantum fluctuations, one needs a sufficient amount of inflation, putting a lower bound on the number of $e$-folds. These conditions combine to set an upper bound on the energy scale of inflation, given by \cite{TCC2}
\begin{eqnarray}\label{H_inf}
\Hinf < 3\sqrt{3} \times 10^{-20} \Mpl\,,
\end{eqnarray}
which can only be realized by a model of low-scale inflation\footnote{This bound can be alleviated if one assumes a non-standard cosmology after the end of inflation \cite{Yun_Long,relax1} or if the equation of state parameter during the bulk of inflation deviates significantly from $w = -1$ \cite{Yun_Long,relax2}. On the other hand, if one takes into account a pre-inflationary dynamics \cite{Piao}, a radiation era leads to an even stronger bound \cite{Yun_Long,Edward}.}. This conclusion is independent of any assumptions on how one obtains inflation. In this note, however, we will be working in the context of the usual assumption that it is a canonically normalized scalar matter field $\varphi$ with potential energy $V(\varphi)$ which is responsible for leading to inflation.

Taking into account the observed scalar power spectrum, one also gets a bound on the slow-roll parameter $\epsilon := \(\Mpl^2/2\)\, \(V'/V\)^2$, such that it is negligibly small \cite{TCC2}
\begin{eqnarray}\label{eps}
\epsilon < 10^{-31}\,.
\end{eqnarray}
If one assumes the standard  single-field consistency relation, it is easy to see that this implies that the primordial tensor-to-scalar ratio ($r$) for inflation must be unobservably tiny \cite{TCC2}
\begin{eqnarray}\label{r}
r < 10^{-30}\,.
\end{eqnarray}
However, since primordial $B$-modes are yet to be observed, this does not necessarily imply doom for inflation. It might be conceivable that, provided one does not invoke alternate mechanisms for the production of primordial tensor modes \cite{Suddho}, a tiny $r$ is part of the predictions of inflationary models. 

Even if a small $r$ does not require one to revisit inflation as the preferred model of early-universe cosmology, there are other potentially troubling consequences of the TCC for inflation. But, first, let us note another interesting observational fact. From the expression for the running of the scalar power spectrum, or the spectral tilt, we know that
\begin{eqnarray}\label{Tilt}
n_s - 1 = 2\eta - 6\epsilon\,,
\end{eqnarray}
where the second slow-roll parameter $\eta := \Mpl^2 \(V''/V\)$. The spectral tilt is quite tightly constrained from observations, and the mean of its measured value from PLANCK is given by $n_s = 0.965$ \cite{Planck}. Since $\epsilon$ is constrained to be negligibly small from the TCC, this implies that 
\begin{eqnarray}\label{eta_obs}
\eta \sim - 0.02\,.
\end{eqnarray}
This conclusion is particularly interesting when one recalls the de-Sitter (dS) swampland conjecture \cite{dSswamp}. The dS conjecture states that for an EFT to \textit{not} be in the swampland\footnote{The {\it swampland} of effective field theories is the set of EFTs which are not consistent with superstring theory - see \cite{Brennan, Palti} for reviews.}, one must have (for $(c, c') \sim \mathcal{O}(1)$)
\begin{eqnarray}\label{dS_conjecture}
\frac{|V'|}{V} > \frac{c}{\Mpl}\;\;\;\;\;\;\; \text{or}\;\;\;\;\;\;\;\; \frac{V''}{V} < \frac{c'}{\Mpl^2}\,,
\end{eqnarray}
$i.e.,$ either the potential is very steep or, when near its maxima, must have large tachyonic instabilities. For inflationary models, this implies that either $\epsilon \sim \mathcal{O}(1)$, or, if $\epsilon \ll 1$, then $|\eta| \sim \mathcal{O}(1)$. Since the TCC implies that $\epsilon \ll 1$, observations \eqref{Tilt} seem to require that $\eta \sim 0.02$, \textit{i.e.}, it is an $\mathcal{O}(1)$ number\footnote{This may be seen as an amelioration of the $eta$-problem since we do not need a value $\eta \ll 1$.}. It might seem that this range of values of $\eta$ might not be enough for avoiding stochastic eternal inflation \cite{Hilltop_EI,No_EI_principle}. On the other hand, several studies have shown that stochastic eternal inflation is in severe conflict with the swampland constraints \cite{No_EI_principle,Swampland_EI}. Note that this apparent contradiction is easily resolved when one also takes into account our required initial value for $\varphi$, as spelled out later in \eqref{phi_i}. The region of parameter space corresponding to this value, as parametrized by \eqref{epsilon} is incompatible with eternal inflation \cite{No_EI_principle}, thus confirming the intuition that the our model is well out of the regime of stochastic eternal inflation.

Given these constraints \eqref{H_inf}, \eqref{eps} and \eqref{Tilt}, models of inflation which can be made viable with the TCC and, more generally, the swampland constraints (see Sec-5), seem to be hilltop ones\footnote{Such potentials can also be more tractable for holographic cosmology \cite{Heliudson}.}. This is so because for hilltop potentials, one generically gets a very tiny $\epsilon$ and a large $\eta$, as is seen to be preferred by the swampland criteria. In the next section, we first give the phenomenological parameters for a hilltop potential given the TCC bounds. However, even allowing for such a model, another potential problem for such a small-field inflationary model lies in the initial condition fine-tuning problem. We shall describe this in Sec-3 before going on to give a solution for it in Sec-4 in the form of the tunnelling wavefunction, which gives a quantum completion for inflation. Finally, in Sec-5, we show how such a hilltop model, and thus the tunnelling wavefunction, is preferred when all of the swampland constraints are taken into account, before concluding in Sec-6.

\section{The Hilltop Potential}

We consider a potential which has the form of a hilltop near $\varphi = 0$, i.e.
\begin{eqnarray}
V(\varphi) = V_0 \[1 - \(\frac{\varphi}{\mu}\)^2\]\, .
\end{eqnarray}
This form of the potential must break down for values of $|\varphi|$ comparable or larger than $\mu$\footnote{In fact, as we will argue later, the breakdown of the potential must occur in fact at much smaller values of $|\varphi|$.}. We will assume that the potential asymptotes to zero for large values of $|\varphi|$ . For this potential, the second slow-roll parameter, for field values $|\varphi| \ll \mu$, takes the form
\begin{eqnarray}\label{eta}
\eta := \Mpl^2 \(\frac{V''}{V}\) \simeq -2 \(\frac{\Mpl}{\mu}\)^2\,.
\end{eqnarray}
Using \eqref{eta_obs}, one gets 
\begin{eqnarray}
\mu \sim 10 \Mpl\,.
\end{eqnarray}
In the above, we have used the fact that $\varphi/\mu \ll 1$ (in \eqref{eta}) which is generically true for a small field excursion model as this one. We shall further justify this later on as well. 

The first slow-roll parameter for this model is given by
\begin{eqnarray}\label{epsilon}
\epsilon := \frac{\Mpl^2}{2} \(\frac{V'}{V}\)^2 \simeq 2 \(\frac{\Mpl}{\mu}\)^2 \(\frac{\varphi}{\mu}\)^2\,,
\end{eqnarray}
where we, once again, use $\varphi/\mu \ll 1$. From the expression of the dimensionless scalar power spectrum,
\begin{eqnarray}
P_s = \frac{1}{8\pi^2\epsilon} \(\frac{H}{\Mpl}\)^2\,,
\end{eqnarray}
one can get the value of $\varphi_i$ at the beginning of the inflationary phase by using \eqref{epsilon} above, as
\begin{eqnarray}\label{phi_i}
\varphi_i \sim 10^5\; \frac{V_0^{1/2}}{\Mpl}\,.
\end{eqnarray}
In the above, we have used $P_s \simeq 2.1 \times 10^{-9}$ and the Friedmann equation 
\begin{eqnarray}
3\Mpl^2 \Hinf^2 \simeq V_0\,.
\end{eqnarray}

Since our goal later on is to show that there exists a mechanism which allows for the quantum mechanical tunnelling of the inflaton to such a  position on field space,  we first need to make sure that such a tunnelling is safe from quantum fluctuations. Note that if indeed the field can tunnel to this position, it is safe from quantum fluctuations displacing it from this position since the amplitude of quantum fluctuations is given by
\begin{eqnarray}
\langle \delta \varphi^2\rangle^{1/2} \simeq \frac{\Hinf}{2\pi} \sim \frac{V_0^{1/2}}{2\sqrt{3}\, \pi \Mpl}\,,
\end{eqnarray}
clearly demonstrating that  $\langle \delta \varphi^2\rangle^{1/2} \ll \varphi_i$. 

Thus, we can fix the two parameters of this model $\mu\sim 10\Mpl$ from observations (and the TCC, implying $\epsilon \ll \eta$), and $V_0^{1/4} < 3 \times 10^{-10}\Mpl$ \cite{TCC2}. The final thing to check is that the classical drift of the field, $\Delta\varphi$ is much smaller than $\mu$, so that one can get enough e-folds of inflation. The TCC implies that the field traverses a very small distance \cite{TCC2}
\begin{eqnarray}
\Delta\varphi &\sim& \sqrt{2 \epsilon}\,\Mpl\, N\,\nonumber\\
&\sim& \frac{V_0^{1/2}}{2\sqrt{3}\,\pi P_s^{1/2} \Mpl}\, N\,.
\end{eqnarray}
Using the bound on the number of $e$-folds imposed by the TCC, $N < \ln\(\Mpl/\Hinf\)$, one gets an estimate for the classical drift as
\begin{eqnarray}
  \Delta \varphi <   10^6\; \frac{V_0^{1/2}}{\Mpl} \ll \mu \, .
\end{eqnarray}
This is the same bound as one got in \cite{TCC2} written in a different way. This shows that $\varphi_i \lesssim \Delta\varphi \ll \mu$ and thus it is possible for inflation to last a sufficient time, if there is a suitable mechanism to begin inflation at $\varphi_i$\footnote{However, one needs to add an additional mechanism to the construction to ensure that inflation does not last too long. The simplest way to achieve this is by a sudden steepening of the potential after the value $\varphi_i + \Delta \varphi$.}.

\section{The Fine-Tuning Problem for Inflation}

There are two \textit{extreme} fine-tuning problems for inflation, as implied by the TCC:
\begin{itemize}
    \item Inflation needs to start near the hilltop, at a value of the scalar field $\varphi_i$, as mentioned in the previous section. From the point of view of classical dynamics, there is no canonical explanation why the field should start at such a small value.
    \item From the point of view of classical dynamical systems, it is natural to assume that the kinetic energy of the inflaton is comparable to the potential energy \textit{before} inflation starts. However, considering the value of $\epsilon$ we require given the TCC and observations, it displays an extreme fine-tuning for the initial velocity of the inflaton as compared to this natural value \cite{TCC2}. This is the usual fine-tuning problem for the initial kinetic energy in small field models of inflation \cite{Dalia}.
\end{itemize}
For large-field inflation models, as is well-known (see e.g. \cite{RHBrev} for a recent review and \cite{Kung} for some initial references), the inflationary slow-roll solution is a local attractor in initial condition space, and the above problems do not show up. However, both these problems appear as the TCC implies that only a small-field model of inflation is allowed, given current bounds on observations. In the next section, we show how the tunnelling wavefunction can solve both of these problems at one go\footnote{In \cite{Kai}, it has been conjectured that some sort of tunnelling might alleviate these problems although the ``tunnelling'' referred in it has nothing to do with the tunnelling wavefunction.}.

In short, the initial condition problems for inflation arise if one considers inflation from the classical physics dynamical systems point of view. However, quantum effects will likely be very important in the early universe. A goal of the field of {\it quantum cosmology} has been to develop a theory for the initial conditions which apply once classical dynamics takes over. According to quantum cosmology (see e.g. \cite{QQ, Halliwell:1990uy} for an overview), the state is described by a {\it wavefunction}, but there are various proposals for how to obtain this wavefunction. The two most popular proposals are the {\it Hartle-Hawking} wavefunction \cite{HH} and the {\it tunnelling wavefunction} \cite{Vilenkin}, although recently potential problems for both approaches have been put forward \cite{Turok}, with several potential resolutions \cite{Rescue}. In the following section, we will study the implications of the {\it tunnelling wavefunction} for the initial condition issue in hilltop models of inflation, making use of its most recent reincarnation \cite{TWF,TWF2}.

\section{The Tunnelling Wavefunction for the Hilltop Potential}

In the case of homogeneous and isotropic cosmology, the tunnelling wavefunction $\Psi$ for Einstein gravity minimally coupled to a canonically normalized scalar field $\varphi$ (evaluated at the values $a_1$ and $\varphi_1$ of the scale factor and scalar field, respectively) is given by a functional integral (following the notation of \cite{TWF}) 
\begin{eqnarray}\label{path}
\Psi_\text{T} \(a_1, \varphi_1\) := \int_0^\infty \d N \int \mathcal{D}[a]\,\mathcal{D}[\varphi] \, e^{i S^{\(0\)} [a,\varphi,N]}\,,
\end{eqnarray}
where $\eta$ is conformal time, $a(\eta)$ is the scale factor, $N(\eta)$ is the lapse function, and
\begin{eqnarray}
 S^{\(0\)} \[a,\varphi,N\] = 6\pi^2 \int_{-\infty}^{\eta_1} \[-\frac{\dot{a}^2}{N} + \frac{a^2 \dot{\varphi}^2}{N} + N \tilde{U}(a, \varphi) \] \d\eta\,,
\end{eqnarray}
is the action. Here, 
\begin{eqnarray}
\tilde{U}(a,\varphi) = a^2 - \frac{a^4\; V\(\varphi\)}{3}\,,
\end{eqnarray}
is the superpotential (we have used $\tilde{U}$ instead of $V$ for the superpotential, as compared to \cite{TWF}, so as not to cause confusion with the scalar potential). In the above, an overdot stands for a derivative with respect to $\eta$. 

It is convenient to choose a time coordinate $t$ \footnote{Note that $t$ is not the usual physical time of homogeneous and isotropic cosmology.}, such that 
\begin{eqnarray}
\d\eta = \frac{1}{a^2} \d t  
\end{eqnarray}
with the line-element taking the form
\begin{eqnarray}
\d s^2 = -\frac{N^2}{q\(t\)} \d t^2 + q\(t\) \d\Omega_3^2\,,
\end{eqnarray}
where $q \equiv a^2$. In these coordinates, the (background) action becomes
\begin{eqnarray}
S^{\(0\)} [q,\varphi,N] = 6\pi^2 \int_0^1 \[-\frac{\dot{q}^2}{4 N} +  \frac{q^2\,\dot{\varphi}^2}{N} + N \(1 -  \frac{q V\(\varphi\)}{3}\) \]\d t\,,
\end{eqnarray}
where the overdots now refer to derivatives with respect to $t$. 

This action is minimized by the solutions of the following classical equations
\begin{eqnarray}\label{EOM}
\ddot{q} - \frac{2}{3} N^2 V\(\varphi\) + 4 q \dot{\varphi}^2 &=& 0\,,\\
\ddot{\varphi} +2 \(\frac{\dot{q}}{q}\) \,\dot{\varphi} +\frac{1}{6} N^2 q \(\frac{\d V}{\d\varphi}\) &=& 0\,.
\end{eqnarray}
We now need to solve this set of equations for our hilltop potential with the parameters described in the Sec-2. 

Our goal is to first solve \eqref{EOM}, with the boundary conditions (some of which are just the regularity conditions for getting a smooth initial wavefunction) for the tunnelling wavefunction
\begin{eqnarray}\label{bc}
q(0) = 0 \;\;\;\; &\text{and}& \;\;\;\; q(1) = a_1^2\\
\dot{\varphi}\(0\) = 0 \;\;\;\;  &\text{and}& \;\;\;\; \varphi(1) =\varphi_1\,.
\end{eqnarray}
For our purposes, it is sufficient to show that the tunnelling wavefunction prefers a large potential energy or, in other words, it would be more probable to tunnel to the hilltop of our given potential. This way we shall be able to explain why the inflaton begins at our required value of $\varphi_i$, due to the measure defined by the tunnelling wavefunction. In this case, it is sufficient to get an analytical solution even if we have to make some approximations to derive it. 

Firstly, we assume the usual slow-roll relation $N^2 V \gg \ 6 q \dot{\varphi}^2$ and the fact that the potential is sufficiently flat. For our form of the potential, this can be easily justified as follows
\begin{eqnarray}
V\(\varphi\) = V_0 \[1 -\(\frac{\varphi}{\mu}\)^2\] &\simeq& V_0 \,,\\
V'\(\varphi\) = - 2 V_0 \(\frac{\varphi}{\mu}\) &\simeq& 0\,,
\end{eqnarray}
which are both valid for $\varphi/\mu \ll 1$, as is the case for our allowed choice of parameter space.

In this case, for our \eqref{bc}, the solutions of \eqref{EOM} take the form
\begin{eqnarray}
q\(t\) &=& \frac{V_0}{3} N^2 t^2 + \(a_1^2 - \frac{V_0}{3} N^2\) t\,,\\
\varphi\(t\) &=& \varphi_1\,.
\end{eqnarray}
The action can now be evaluated for this soluion and yields
\begin{eqnarray}
S^{\(0\)} [a_1,\varphi_1, N] \simeq 6\pi^2 \[\frac{N^3 V^2}{108} + N\(1- \frac{V}{6}  a_1^2\) - \frac{a_1^4}{N}\]\,,
\end{eqnarray}
with 
\begin{eqnarray}
V \simeq V_0\[1 - \(\frac{\varphi_1}{\mu}\)^2\]\,. 
\end{eqnarray}
We can now perform the lapse integration using the saddle point approximation. The saddles for the lapse are given by the following equation:
\begin{eqnarray}
V^2 N_s^4 + \(36 - 6 V a_1^2\) N_s + 9 a_1^4 = 0\,.
\end{eqnarray}

We are interested in the values of $N$ in the underbarrier regions which are given by the  solutions
\begin{eqnarray}
N^{\pm} = \frac{3\, i}{V} \(1 \mp \sqrt{1 - \frac{V}{3} a_1^2} \)\,,
\end{eqnarray}
with the corresponding values of the saddle-point action
\begin{eqnarray}
S^{\(0\)} \[a_1, \varphi_1, N^{\pm}\] = 12\pi^2 i S^{\pm} \[a_1, \varphi_1\]\,,
\end{eqnarray}
with
\begin{eqnarray}
S^{\pm} \[a_1, \varphi_1\] := \mp \int_{a_1}^{a_*} \sqrt{U\(a',\varphi_1\)}\; \d a' + \int_0^{a_*} \sqrt{U\(a',\varphi_1\)}\; \d a'\,,
\end{eqnarray}
where $U = a^2 \tilde{U}$. Here $a_*$ is defined as the turning point where $U\(a_*, \varphi_1\) = 0$. The important thing for us is that under our given approximations, $\varphi$ is a constant, and the tunnelling wavefunction is essentially the same as that for the cosmological constant case. In this case, the amplitude of nucleation is given by the solution for the only saddle-point which contributes to the classically allowed regime $U\(a_1, \varphi_1\) < 0$,
\begin{eqnarray}
N = \frac{3}{V} \(i + \sqrt{\frac{V}{3}a_1^2 - 1}\)\, .
\end{eqnarray}

The probability of nucleation is given by
\begin{eqnarray}
\mathcal{P} \propto \exp{\(-12\pi^2 S\(a_1, \varphi_1\)\)}\,,
\end{eqnarray}
with
\begin{eqnarray}
S(a_1) = i\int_{a_*}^{a_1} \sqrt{-U\(a', \varphi_1\)}\, \d a'  + \int_0^{a_*} \sqrt{U\(a',\varphi_1\)}\; \d a'\,.
\end{eqnarray}
The semiclassical factor contributing to the probability comes from the second term above. After a little algebra, one can see that the probability of nucleation for the tunnelling wavefunction is given by
\begin{eqnarray}
	\mathcal{P} \propto e^ {-1/V}\,,
\end{eqnarray}
ignoring some numerical factors. This is the same probability one gets for the tunnelling wavefunction by solving the Wheeler-de Witt equation with the tunnelling boundary conditions (in the canonical formulation as opposed to the path integral version followed here).

This result clearly demonstrates that the tunnelling wave function prefers a nucleation to the top of the hilltop since the probability decreases sharply as the potential energy gets smaller. The exact top of the hill occurs at $\varphi_\text{top} \equiv 0$. We are interested in the probability that the field nucleates at the specific value of $\varphi \sim \varphi_i$. However, given the value of $\mu$ which we are using, $\varphi_i$ is close to the top of the hill in terms of the value of the potential energy. If we compare the probability to nucleate at a value $\varphi \sim \varphi_i$ to its maximum value, we have
\begin{eqnarray}
\log \(\frac{\mathcal{P}}{\mathcal{P}_{\mathrm{max}}}\) \propto - \frac{1}{V(\varphi)} + \frac{1}{V_{0}} \simeq - \frac{\varphi^{2}}{V_{0} \mu^{2}},
\end{eqnarray}
where this approximation is true if $\varphi/\mu \ll 1$. Therefore, the probability difference between nucleating near $\varphi_i$ and $\varphi_\text{top}$ for the hilltop potential is negligibly small. Since, in order not to obtain too long a period of inflation, the value of the potential has to decrease sharply beyond the value of $\varphi_i + \Delta \varphi$, the tunnelling probability to a value of $|\varphi|$ larger than that one is negligible. Hence, the probability of nucleating with $|\varphi| \sim \varphi_i$ is of the order
\begin{equation}
{\cal{P}} \, \sim \, \frac{\varphi_i}{\varphi_i + \Delta \varphi} \, ,
\end{equation}
which is of the order $10^{-1}$ given the numbers we have used in Sec-2. We thus see that the tunnelling wavefunction gives a  simple explanation why inflation should begin at $\varphi_i$ for the form of the potential which we are using.

The second fine-tuning requirement for inflation from the TCC is that the initial field velocity for the inflation is extremely small. However, for the tunnelling wavefunction, this condition is automatically realized since the velocity at the ``South Pole'' of the Euclidean instanton has to be precisely zero due to regularity. In other words, we can see from \eqref{bc} that $\dot\varphi(0) =0$ and for a constant potential, we get the ``no-roll'' solution. However, once we introduce our (slow-roll) hilltop potential, there would be a very small velocity for the inflaton after it tunnels to the hilltop. Nevertheless, this velocity would be extremely tiny and can naturally explain the slow-velocity required for the inflaton given the TCC. In other words, following the logic of \cite{TCC2}, we provide a quantum mechanical explanation for why the velocity of the scalar field \textit{should not} be large before inflation begins. In fact, for our case, it should be almost zero before inflation starts.

\section{Swampland and the Tunnelling Wavefunction}

Having established how the tunnelling wavefunction can help us attain the initial condition, as required by the TCC, let us point out how the same arguments would also work more generally for the other swampland conjectures. Firstly, recall that the swampland distance conjecture (SDC) \cite{SDC} states that the field excursion during inflation should be less than some $\mathcal{O}(1)$ number in Planck units for the EFT to be under control. This is easily achieved by the low-scale hilltop model described in this paper. 

Secondly, the dS conjecture tells us that either the slope of the potential or its second derivative must be large, as quantified in \eqref{dS_conjecture}. Unless we invoke additional fields or other degrees of freedom in the form of a modified initial state or an action deviating from GR, it is not possible to have a large slope for an inflationary potential due to the tight observational bound on the upper limit of $r$. In other words, for single-field\footnote{To be more precise, we should use the terminology single-clock in this context.}, slow-roll models of inflation, one cannot satisfy the condition for a large slope (or, equivalently a large $\epsilon$). For exceptions, see \cite{Swampland_Inflation}. On the other hand, it is quite possible to allow for models with a large $\eta$ provided $\epsilon$ remains small. The prototype for such an inflationary potential is indeed the hilltop model with more complicated versions allowing for additional terms in higher polynomial powers of the inflaton. Indeed, in all examples of such inflationary models (e.g., natural inflation \cite{natural} with a cosine term) the potential can be expanded in a power series which looks like our simple hilltop potential near the maxima. Therefore, if one is to look for a model of single-field, slow-roll inflation, in the spirit of finding the simplest EFT of a minimally-coupled scalar to GR with some potential, the hilltop model is the preferred choice in view of all the swampland conjectures. It is not surprising that both the TCC as well as the SDC and dS conjecture all prefer the same type of scalar potential, given the fact that these different swampland conjectures have been shown to be related (and, indeed, do follow) to each other \cite{TCC, Suddho2}.

Given the somewhat special status of the hilltop potential in inflationary model-building due to these theoretical considerations, it is now clear why the tunnelling wavefunction becomes important in setting initial conditions for inflation. In this work, we have demonstrated how the tunnelling wavefunction does an excellent job of explaining why in hilltop models an initial value of the inflaton close to the top of the hill should be preferred as the starting point in the phase of classical evolution with a negligibly small velocity. Keeping in mind that this is a truly quantum cosmological boundary condition gives us hope that there is a deeper reason to explore such proposals for quantum gravity more seriously in our quest for deriving dS spacetime as a low-energy EFT. We have to add, however, that from the point of view of an EFT it is unnatural to obtain the fairly sharp cutoff of the potential at a value $\varphi \sim \varphi_i + \Delta \varphi$ which must be added (to make sure that the TCC remains satisfied), and this form of the potential needs to be justified from fundamental string theory for satisfactory model-building.

\section{Conclusion}

In this note we have argued that the tunnelling wavefunction can provide, without too much tuning, the initial conditions required for hilltop inflation modes which are consistent with the TCC, initial conditions which from the point of view of a classical dynamical systems approach look highly fine tuned. The hilltop potentials we use are consistent with the swampland conjectures on effective field theories which can emerge from superstring theory. On the other hand, rendering a hilltop inflation model consistent with the short duration of inflation which the TCC only allows requires introducing a sharp cutoff in the potential energy function which does not look natural.

Finally, it is worth pointing out that the conclusions reached from the tunnelling wavefunction do not naturally follow from other quantum gravity proposals such as the no-boundary wavefunction. In fact, for the minimal model of a single-field slow-roll potential, the no-boundary wavefunction would prefer tunnelling to the local minima \cite{HHH} (and certainly not to the top of the hillt). Given the swampland constraints, this shows us that the tunnelling proposal indeed occupies a special place as the preferred boundary condition for the wavefunction of the universe.

As this manuscript was being finalized for submission, a paper \cite{Tenkanen} appeared which also discusses the initial conditions for plateau models of inflation, but from a very different point of view.

\section*{Acknowledgements}
This research is supported in part by funds from NSERC, from the Canada Research Chair program, by a McGill Space Institute fellowship and by a generous gift from John Greig. DY is supported by the National Research Foundation of Korea (grant no. 2018R1D1A1B07049126).


\begin{thebibliography}{99}

\bibitem{TCC}
A.~Bedroya and C.~Vafa,
  ``Trans-Planckian Censorship and the Swampland,''
  arXiv:1909.11063 [hep-th].
  %%CITATION = ARXIV:1909.11063;%%    
  
\bibitem{ChibMukh}
V. Mukhanov and G. Chibisov,
 ``Quantum Fluctuation And Nonsingular Universe. (In Russian),''
 JETP Lett.\  {\bf 33}, 532 (1981) [Pisma Zh.\ Eksp.\ Teor.\ Fiz.\  {\bf 33}, 549 (1981)].
 %%CITATION = ZFPRA,33,549;%%
 
\bibitem{Starob}
A.~A.~Starobinsky,
``Spectrum of relict gravitational radiation and the early state of the universe,''
  JETP Lett.\  {\bf 30}, 682 (1979)
  [Pisma Zh.\ Eksp.\ Teor.\ Fiz.\  {\bf 30}, 719 (1979)].
  %%CITATION = JTPLA,30,682;%%
  
\bibitem{Jerome}
J.~Martin and R.~H.~Brandenberger,
  ``The TransPlanckian problem of inflationary cosmology,''
  Phys.\ Rev.\ D {\bf 63}, 123501 (2001)
  [hep-th/0005209];\\
  %%CITATION = doi:10.1103/PhysRevD.63.123501;%%
  R.~H.~Brandenberger and J.~Martin,
  ``Trans-Planckian Issues for Inflationary Cosmology,''
  Class.\ Quant.\ Grav.\  {\bf 30}, 113001 (2013)
  [arXiv:1211.6753 [astro-ph.CO]].
  %%CITATION = doi:10.1088/0264-9381/30/11/113001;%%    
  
\bibitem{TCC2} 
A.~Bedroya, R.~Brandenberger, M.~Loverde and C.~Vafa,
``Trans-Planckian Censorship and Inflationary Cosmology,''
arXiv:1909.11106 [hep-th].

\bibitem{Yun_Long}
S.~Mizuno, S.~Mukohyama, S.~Pi and Y.~L.~Zhang,
  ``Universal Upper Bound on the Inflationary Energy Scale from the Trans-Planckian Censorship Conjecture,''
  arXiv:1910.02979 [astro-ph.CO].
  
  
\bibitem{relax1}  
  M.~Dhuria and G.~Goswami,
  ``Trans-Planckian Censorship Conjecture and Non-thermal post-inflationary history,''
  arXiv:1910.06233 [astro-ph.CO];\\
  %%CITATION = ARXIV:1910.06233;%%
  M.~Torabian,
  ``Non-Standard Cosmological Models and the trans-Planckian Censorship Conjecture,''
  arXiv:1910.06867 [hep-th];\\
  %%CITATION = ARXIV:1910.06867;%%
 H.~H.~Li, G.~Ye, Y.~Cai and Y.~S.~Piao,
  ``Trans-Planckian censorship of multi-stage inflation and dark energy,''
  arXiv:1911.06148 [gr-qc];\\
  %%CITATION = ARXIV:1911.06148;%% 
M.~Torabian,
  ``Breathing Comoving Hubble: Initial Condition and Eternity in view of the trans-Planckian Censorship Conjecture,''
  arXiv:1911.12304 [hep-th].
  %%CITATION = ARXIV:1911.12304;%% 

\bibitem{relax2}
V. Kamali and R. Brandenberger
 ``Relaxing the TCC Bound on Inflationary Cosmology? ``
 arXiv:2001.00040;\\
  %%CITATION = ARXIV:2001.00040;%%
  A.~Berera and J.~R.~Calderón,
  ``Trans-Planckian censorship and other swampland bothers addressed in warm inflation,''
  Phys.\ Rev.\ D {\bf 100}, no. 12, 123530 (2019)
  arXiv:1910.10516 [hep-ph].

\bibitem{Piao}
Y.~Cai and Y.~S.~Piao,
``Pre-inflation and Trans-Planckian Censorship,''
arXiv:1909.12719 [gr-qc].

  
\bibitem{Edward}
R. Brandenberger and E. Wilson-Ewing,
``Strengthening the TCC Bound on Inflationary Cosmology'',
arXiv:2001.00043 [hep-th].
%%CITATION = ARXIV:2001.00043;%%



\bibitem{Suddho} 
S.~Brahma,
``Trans-Planckian censorship, inflation and excited initial states for perturbations,''
Phys.\ Rev.\ D {\bf 101}, no. 2, 023526 (2020)
% doi:10.1103/PhysRevD.101.023526
arXiv:1910.04741 [hep-th].

\bibitem{Planck}
N.~Aghanim {\it et al.} [Planck Collaboration],
  ``Planck 2018 results. VI. Cosmological parameters,''
  arXiv:1807.06209 [astro-ph.CO].
  %%CITATION = ARXIV:1807.06209;%%
  
\bibitem{dSswamp}
 G.~Obied, H.~Ooguri, L.~Spodyneiko and C.~Vafa,
 ``De Sitter Space and the Swampland,''
 arXiv:1806.08362 [hep-th];\\
 %%CITATION = ARXIV:1806.08362;%%
 H.~Ooguri, E.~Palti, G.~Shiu and C.~Vafa,
  ``Distance and de Sitter Conjectures on the Swampland,''
  Phys.\ Lett.\ B {\bf 788}, 180 (2019)
  [arXiv:1810.05506 [hep-th]];\\
  %%CITATION = doi:10.1016/j.physletb.2018.11.018;%%
  S.~K.~Garg and C.~Krishnan,
  ``Bounds on Slow Roll and the de Sitter Swampland,''
  JHEP {\bf 1911}, 075 (2019)
  %doi:10.1007/JHEP11(2019)075
  [arXiv:1807.05193 [hep-th]];\\
  %%CITATION = doi:10.1007/JHEP11(2019)075;%%
  D.~Andriot and C.~Roupec,
  %``Further refining the de Sitter swampland conjecture,''
  Fortsch.\ Phys.\  {\bf 67}, no. 1-2, 1800105 (2019)
  arXiv:1811.08889 [hep-th].
  
\bibitem{Brennan}
T.~D.~Brennan, F.~Carta and C.~Vafa,
 ``The String Landscape, the Swampland, and the Missing Corner,''
 PoS TASI {\bf 2017}, 015 (2017)
 %doi:10.22323/1.305.0015
arXiv:1711.00864 [hep-th].
 %%CITATION = doi:10.22323/1.305.0015;%%
 
\bibitem{Palti}
E.~Palti,
 ``The Swampland: Introduction and Review,''
 arXiv:1903.06239 [hep-th].
 %%CITATION = ARXIV:1903.06239;%%   

\bibitem{Hilltop_EI}
 W.~H.~Kinney,
``Eternal Inflation and the Refined Swampland Conjecture,''
Phys.\ Rev.\ Lett.\  {\bf 122}, no. 8, 081302 (2019)
arXiv:1811.11698 [astro-ph.CO].

\bibitem{No_EI_principle}
T.~Rudelius,
``Conditions for (No) Eternal Inflation,''
JCAP {\bf 1908}, 009 (2019)
arXiv:1905.05198 [hep-th].

\bibitem{Swampland_EI}
S.~Brahma and S.~Shandera,
``Stochastic eternal inflation is in the swampland,''
JHEP {\bf 1911}, 016 (2019)
arXiv:1904.10979 [hep-th];\\
Z.~Wang, R.~Brandenberger and L.~Heisenberg,
``Eternal Inflation, Entropy Bounds and the Swampland,''
arXiv:1907.08943 [hep-th].
 
\bibitem{Heliudson} 
H.~Bernardo,
``On the TransPlanckian Censorship Conjecture in Holographic Cosmology,''
arXiv:1912.00100 [hep-th].

\bibitem{Dalia}
D.~S.~Goldwirth and T.~Piran,
  ``Initial conditions for inflation,''
  Phys.\ Rept.\  {\bf 214}, 223 (1992).
  %doi:10.1016/0370-1573(92)90073-9
  %%CITATION = doi:10.1016/0370-1573(92)90073-9;%%
  
\bibitem{RHBrev}
R.~Brandenberger,
  ``Initial conditions for inflation — A short review,''
  Int.\ J.\ Mod.\ Phys.\ D {\bf 26}, no. 01, 1740002 (2016)
  %doi:10.1142/S0218271817400028
  arXiv:1601.01918 [hep-th].
  %%CITATION = doi:10.1142/S0218271817400028;%%
  
\bibitem{Kung}
A.~Albrecht, R.~H.~Brandenberger and R.~Matzner,
  ``Inflation With Generalized Initial Conditions,''
  Phys.\ Rev.\ D {\bf 35}, 429 (1987);\\
 % doi:10.1103/PhysRevD.35.429;\\
  %%CITATION = doi:10.1103/PhysRevD.35.429;%%
H.~Kurki-Suonio, R.~A.~Matzner, J.~Centrella and J.~R.~Wilson,
  ``Inflation From Inhomogeneous Initial Data in a One-dimensional Back Reacting Cosmology,''
  Phys.\ Rev.\ D {\bf 35}, 435 (1987);\\
  %doi:10.1103/PhysRevD.35.435;\\
  %%CITATION = doi:10.1103/PhysRevD.35.435;%%
R.~H.~Brandenberger and J.~H.~Kung,
  ``Chaotic Inflation as an Attractor in Initial Condition Space,''
  Phys.\ Rev.\ D {\bf 42}, 1008 (1990);\\
  %doi:10.1103/PhysRevD.42.1008;\\
  %%CITATION = doi:10.1103/PhysRevD.42.1008;%%
H.~A.~Feldman and R.~H.~Brandenberger,
  ``Chaotic Inflation With Metric and Matter Perturbations,''
  Phys.\ Lett.\ B {\bf 227}, 359 (1989);\\
  %doi:10.1016/0370-2693(89)90944-1;\\
  %%CITATION = doi:10.1016/0370-2693(89)90944-1;%%   
W.~E.~East, M.~Kleban, A.~Linde and L.~Senatore,
  ``Beginning inflation in an inhomogeneous universe,''
  JCAP {\bf 1609}, 010 (2016)
  %doi:10.1088/1475-7516/2016/09/010
  arXiv:1511.05143 [hep-th];\\
  %%CITATION = doi:10.1088/1475-7516/2016/09/010;%%      
  K.~Clough, E.~A.~Lim, B.~S.~DiNunno, W.~Fischler, R.~Flauger and S.~Paban,
  ``Robustness of Inflation to Inhomogeneous Initial Conditions,''
  JCAP {\bf 1709}, 025 (2017)
  %doi:10.1088/1475-7516/2017/09/025
  arXiv:1608.04408 [hep-th].
  %%CITATION = doi:10.1088/1475-7516/2017/09/025;%%  

\bibitem{Kai}
K.~Schmitz,
``Trans-Planckian Censorship and Inflation in Grand Unified Theories,''
arXiv:1910.08837 [hep-ph].

\bibitem{QQ}
 J.~B.~Hartle,
  ``Space-time quantum mechanics and the quantum mechanics of space-time,''
  gr-qc/9304006.
  %%CITATION = GR-QC/9304006;%% 

\bibitem{Halliwell:1990uy} 
J.~J.~Halliwell,
``Introductory Lectures On Quantum Cosmology,''
arXiv:0909.2566 [gr-qc].

\bibitem{HH}
J.~B.~Hartle and S.~W.~Hawking,
  ``Wave Function of the Universe,''
  Phys.\ Rev.\ D {\bf 28}, 2960 (1983)
  [Adv.\ Ser.\ Astrophys.\ Cosmol.\  {\bf 3}, 174 (1987)].
%  doi:10.1103/PhysRevD.28.2960
  %%CITATION = doi:10.1103/PhysRevD.28.2960;%%
  
\bibitem{Vilenkin}
A.~Vilenkin,
  ``Creation of Universes from Nothing,''
  Phys.\ Lett.\  {\bf 117B}, 25 (1982).
 % doi:10.1016/0370-2693(82)90866-8;\\
  %%CITATION = doi:10.1016/0370-2693(82)90866-8;%%
A.~Vilenkin,
  ``Quantum Creation of Universes,''
  Phys.\ Rev.\ D {\bf 30}, 509 (1984).
  %doi:10.1103/PhysRevD.30.509
  %%CITATION = doi:10.1103/PhysRevD.30.509;%%
  
\bibitem{Turok}
 J.~Feldbrugge, J.~L.~Lehners and N.~Turok,
  ``Lorentzian Quantum Cosmology,''
  Phys.\ Rev.\ D {\bf 95}, no. 10, 103508 (2017)
  %doi:10.1103/PhysRevD.95.103508
  arXiv:1703.02076 [hep-th];\\
  %%CITATION = doi:10.1103/PhysRevD.95.103508;%%  
   J.~Feldbrugge, J.~L.~Lehners and N.~Turok,
  ``No smooth beginning for spacetime,''
  Phys.\ Rev.\ Lett.\  {\bf 119}, no. 17, 171301 (2017)
  arXiv:1705.00192 [hep-th];\\
   J.~Feldbrugge, J.~L.~Lehners and N.~Turok,
  ``No rescue for the no boundary proposal: Pointers to the future of quantum cosmology,''
  Phys.\ Rev.\ D {\bf 97}, no. 2, 023509 (2018)
   arXiv:1708.05104 [hep-th];\\
   A.~Di Tucci, J.~Feldbrugge, J.~L.~Lehners and N.~Turok,
   ``Quantum Incompleteness of Inflation,''
   Phys.\ Rev.\ D {\bf 100}, no. 6, 063517 (2019)
   arXiv:1906.09007 [hep-th].
  

\bibitem{Rescue}
J.~Diaz Dorronsoro, J.~J.~Halliwell, J.~B.~Hartle, T.~Hertog, O.~Janssen and Y.~Vreys,
``Damped perturbations in the no-boundary state,''
Phys.\ Rev.\ Lett.\  {\bf 121} (2018) no.8,  081302
arXiv:1804.01102 [gr-qc];\\
M.~Bojowald and S.~Brahma,
``Loops rescue the no-boundary proposal,''
Phys.\ Rev.\ Lett.\  {\bf 121}, no. 20, 201301 (2018)
arXiv:1810.09871 [gr-qc];\\
A.~Di Tucci and J.~L.~Lehners,
``No-Boundary Proposal as a Path Integral with Robin Boundary Conditions,''
Phys.\ Rev.\ Lett.\  {\bf 122}, no. 20, 201302 (2019)
arXiv:1903.06757 [hep-th];\\
O.~Janssen, J.~J.~Halliwell and T.~Hertog,
``No-boundary proposal in biaxial Bianchi IX minisuperspace,''
Phys.\ Rev.\ D {\bf 99}, no. 12, 123531 (2019)
arXiv:1904.11602 [gr-qc].
           
\bibitem{TWF} 
  A.~Vilenkin and M.~Yamada,
  ``Tunneling wave function of the universe,''
  Phys.\ Rev.\ D {\bf 98}, no. 6, 066003 (2018)
 %doi:10.1103/PhysRevD.98.066003
  arXiv:1808.02032 [gr-qc].
  
\bibitem{TWF2}  
   A.~Vilenkin and M.~Yamada,
  ``Tunneling wave function of the universe II: the backreaction problem,''
  Phys.\ Rev.\ D {\bf 99}, no. 6, 066010 (2019)
 arXiv:1812.08084 [gr-qc].

\bibitem{SDC}  
H.~Ooguri and C.~Vafa,
 ``On the Geometry of the String Landscape and the Swampland,''
 Nucl.\ Phys.\ B {\bf 766}, 21 (2007)
 %doi:10.1016/j.nuclphysb.2006.10.033
 [hep-th/0605264].
 %%CITATION = doi:10.1016/j.nuclphysb.2006.10.033;%%

\bibitem{Swampland_Inflation}
W.~H.~Kinney, S.~Vagnozzi and L.~Visinelli,
``The zoo plot meets the swampland: mutual (in)consistency of single-field inflation, string conjectures, and cosmological data,''
Class.\ Quant.\ Grav.\  {\bf 36}, no. 11, 117001 (2019)
arXiv:1808.06424 [astro-ph.CO];\\
S.~Brahma and M.~Wali Hossain,
``Avoiding the string swampland in single-field inflation: Excited initial states,''
JHEP {\bf 1903}, 006 (2019)
arXiv:1809.01277 [hep-th];\\
H.~Geng,
``A Potential Mechanism for Inflation from Swampland Conjectures,''
arXiv:1910.14047 [hep-th].


 
 \bibitem{natural}
 K.~Freese, J.~A.~Frieman and A.~V.~Olinto,
  ``Natural inflation with pseudo - Nambu-Goldstone bosons,''
  Phys.\ Rev.\ Lett.\  {\bf 65}, 3233 (1990).
  %doi:10.1103/PhysRevLett.65.3233
  %%CITATION = doi:10.1103/PhysRevLett.65.3233;%%
  
 \bibitem{Suddho2}
S.~Brahma,
  ``The trans-Planckian censorship conjecture from the swampland distance conjecture,''
  arXiv:1910.12352 [hep-th].
  %%CITATION = ARXIV:1910.12352;%% 
  
\bibitem{HHH} 
J.~B.~Hartle, S.~W.~Hawking and T.~Hertog,
``No-Boundary Measure of the Universe,''
Phys.\ Rev.\ Lett.\  {\bf 100}, 201301 (2008)
arXiv:0711.4630 [hep-th];\\
J.~B.~Hartle, S.~W.~Hawking and T.~Hertog,
``The Classical Universes of the No-Boundary Quantum State,''
Phys.\ Rev.\ D {\bf 77}, 123537 (2008)
arXiv:0803.1663 [hep-th].
  
\bibitem{Tenkanen}
 T. Tenkanen and E. Tomberg,
 ``Initial Conditions for Plateau Inflation'',
 arXiv:2002.02420.
 
\end{thebibliography}
\end{document}